\documentclass[twocolumn,showpacs,prl,superscriptaddress]{revtex4}
\usepackage{epsfig}
\begin{document}

\title{First results from the CERN Axion Solar Telescope (CAST)}

\newcommand{\CERN}{European Organization for Nuclear Research (CERN), Gen\`eve, Switzerland}
\newcommand{\Saclay}{DAPNIA, Centre d'\'Etudes Nucl\'eaires de Saclay (CEA-Saclay), Gif-sur-Yvette, France}
\newcommand{\SCarolina}{Department of Physics and Astronomy, University of South Carolina, Columbia, SC, USA}
\newcommand{\Darmstadt}{GSI-Darmstadt and Institut f\"{u}r Kernphysik, TU Darmstadt, Darmstadt, Germany}
\newcommand{\MPE}{Max-Planck-Institut f\"{u}r Extraterrestrische Physik, Garching, Germany}
\newcommand{\Zaragoza}{Instituto de F\'{\i}sica Nuclear y Altas Energ\'{\i}as, Universidad de Zaragoza, Zaragoza, Spain }
\newcommand{\Chicago}{Enrico Fermi Institute and KICP, University of Chicago, Chicago, IL, USA}
\newcommand{\Thessaloniki}{Aristotle University of Thessaloniki, Thessaloniki, Greece}
\newcommand{\Athens}{National Center for Scientific Research ``Demokritos'', Athens, Greece}
\newcommand{\Freiburg}{Albert-Ludwigs-Universit\"{a}t Freiburg, Freiburg, Germany}
\newcommand{\INR}{Institute for Nuclear Research (INR), Russian Academy of Sciences, Moscow, Russia}
\newcommand{\Vancouver}{Department of Physics and Astronomy, University of British Columbia, Vancouver, Canada }
\newcommand{\Frankfurt}{Johann Wolfgang Goethe-Universit\"at, Institut f\"ur Angewandte Physik, Frankfurt am Main, Germany}
\newcommand{\MPI}{Max-Planck-Institut f\"{u}r Physik (Werner-Heisenberg-Institut), Munich, Germany}
\newcommand{\Zagreb}{Rudjer Bo\v{s}kovi\'{c} Institute, Zagreb, Croatia}

\newcommand{\Pisa}{Scuola Normale Superiore, Pisa, Italy.}
\newcommand{\Lyon}{Inst. de Physique Nucl\'eaire, Lyon, France.}
\newcommand{\BNL}{Brookhaven Nat. Lab., NY-USA.}

\affiliation{\CERN}
\affiliation{\Saclay}
\affiliation{\SCarolina}
\affiliation{\Darmstadt}
\affiliation{\MPE}
\affiliation{\Zaragoza}
\affiliation{\Chicago}
\affiliation{\Thessaloniki}
\affiliation{\Athens}
\affiliation{\Freiburg}
\affiliation{\INR}
\affiliation{\Vancouver}
\affiliation{\Frankfurt}
\affiliation{\MPI}
\affiliation{\Zagreb}

\author{    K.~Zioutas  }\affiliation{\Thessaloniki}
\author{    S.~Andriamonje  }\affiliation{\Saclay}
\author{    V.~Arsov  }\affiliation{\Frankfurt}\affiliation{\Darmstadt}
\author{    S.~Aune  }\affiliation{\Saclay}
\author{    D.~Autiero  }\altaffiliation[Present addr.: ]{\Lyon}\affiliation{\CERN}
\author{    F.~Avignone  }\affiliation{\SCarolina}
\author{    K.~Barth  }\affiliation{\CERN}
\author{    A.~Belov  }\affiliation{\INR}
\author{    B.~Beltr\'an  }\affiliation{\Zaragoza}
\author{    H.~Br\"auninger  }\affiliation{\MPE}
\author{    J.~M.~Carmona  }\affiliation{\Zaragoza}
\author{    S.~Cebri\'an  }\affiliation{\Zaragoza}
\author{    E.~Chesi  }\affiliation{\CERN}
\author{    J.~I.~Collar  }\affiliation{\Chicago}
\author{    R.~Creswick  }\affiliation{\SCarolina}
\author{    T.~Dafni  }\affiliation{\Darmstadt}
\author{    M.~Davenport  }\affiliation{\CERN}
\author{    L.~Di~Lella  }\altaffiliation[Present addr.: ]{\Pisa}\affiliation{\CERN}
\author{    C.~Eleftheriadis  }\affiliation{\Thessaloniki}
\author{    J.~Englhauser  }\affiliation{\MPE}
\author{    G.~Fanourakis  }\affiliation{\Athens}
\author{    H.~Farach  }\affiliation{\SCarolina}
\author{    E.~Ferrer  }\affiliation{\Saclay}
\author{    H.~Fischer  }\affiliation{\Freiburg}
\author{    J.~Franz  }\affiliation{\Freiburg}
\author{    P.~Friedrich    }\affiliation{\MPE}
\author{    T.~Geralis  }\affiliation{\Athens}
\author{    I.~Giomataris  }\affiliation{\Saclay}
\author{    S.~Gninenko  }\affiliation{\INR}
\author{    N.~Goloubev  }\affiliation{\INR}
\author{    M.~D.~Hasinoff  }\affiliation{\Vancouver}
\author{    F.~H.~Heinsius  }\affiliation{\Freiburg}
\author{    D.H.H.~Hoffmann  }\affiliation{\Darmstadt}
\author{    I.~G.~Irastorza  }\affiliation{\Saclay}
\author{    J.~Jacoby  }\affiliation{\Frankfurt}
\author{    D.~Kang  }\affiliation{\Freiburg}
\author{    K.~K\"onigsmann  }\affiliation{\Freiburg}
\author{    R.~Kotthaus  }\affiliation{\MPI}
\author{    M.~Kr\v{c}mar  }\affiliation{\Zagreb}
\author{    K.~Kousouris  }\affiliation{\Athens}
\author{    M.~Kuster  }\affiliation{\MPE}
\author{    B.~Laki\'{c}  }\affiliation{\Zagreb}
\author{    C.~Lasseur  }\affiliation{\CERN}
\author{    A.~Liolios  }\affiliation{\Thessaloniki}
\author{    A.~Ljubi\v{c}i\'{c}  }\affiliation{\Zagreb}
\author{    G.~Lutz  }\affiliation{\MPI}
\author{    G.~Luz\'on  }\affiliation{\Zaragoza}
\author{    D.~W.~Miller  }\affiliation{\Chicago}
\author{    A.~Morales  }\altaffiliation[Deceased.]{}\affiliation{\Zaragoza}
\author{    J.~Morales  }\affiliation{\Zaragoza}
\author{    M.~Mutterer  }\affiliation{\Darmstadt}
\author{    A.~Nikolaidis  }\affiliation{\Thessaloniki}
\author{    A.~Ortiz  }\affiliation{\Zaragoza}
\author{    T.~Papaevangelou  }\affiliation{\CERN}
\author{    A.~Placci  }\affiliation{\CERN}
\author{    G.~Raffelt  }\affiliation{\MPI}
\author{    J.~Ruz  }\affiliation{\Zaragoza}
\author{    H.~Riege  }\affiliation{\Darmstadt}
\author{    M.~L.~Sarsa  }\affiliation{\Zaragoza}
\author{    I.~Savvidis  }\affiliation{\Thessaloniki}
\author{    W.~Serber  }\affiliation{\MPI}
\author{    P.~Serpico  }\affiliation{\MPI}
\author{    Y.~Semertzidis   }\altaffiliation[Permanent addr.: ]{\BNL}\affiliation{\Darmstadt}
\author{    L.~Stewart  }\affiliation{\CERN}
\author{    J.~D.~Vieira  }\affiliation{\Chicago}
\author{    J.~Villar  }\affiliation{\Zaragoza}
\author{    L.~Walckiers  }\affiliation{\CERN}
\author{    K.~Zachariadou  }\affiliation{\Athens}

\collaboration{CAST Collaboration} \noaffiliation

\date{\today}

\begin{abstract}
Hypothetical axion-like particles with a two-photon interaction
would be produced in the Sun by the Primakoff process.  In a
laboratory magnetic field (``axion helioscope'') they would be
transformed into X-rays with energies of a few keV.  Using a
decommissioned LHC test magnet, CAST ran for about 6 months during
2003. The first results from the analysis of these data are
presented here. No signal above background was observed, implying
an upper limit to the axion-photon coupling $g_{a\gamma} < 1.16
\times 10^{-10}~{\rm GeV}^{-1}$ at 95\% CL for $m_a\alt 0.02~{\rm
eV}$. This limit, assumption-free, is comparable to the limit from
stellar energy-loss arguments and considerably more restrictive
than any previous experiment over a broad range of axion masses.
\end{abstract}

\maketitle
%%%%%%%%%%%%%%%%%%%%%%%%%%%%%%%%%%%%%%%%%%%%%%%%%%%%%%%%%%%%%%%%%%%%%%
%% Introduction %%%%%%%%%%%%%%%%%%%%%%%%%%%%%%%%%%%%%%%%%%%%%%%%%%%%%%
%%%%%%%%%%%%%%%%%%%%%%%%%%%%%%%%%%%%%%%%%%%%%%%%%%%%%%%%%%%%%%%%%%%%%%
{\em Introduction.}---Neutral pions, gravitons, hypothetical
axions, or other particles with a two-photon interaction can
transform into photons in external electric or magnetic fields, an
effect first discussed by Primakoff in the early days of pion
physics~\cite{Primakoff}. Therefore, stars could produce these
particles by transforming thermal photons in the fluctuating
electromagnetic fields of the stellar
plasma~\cite{Dicus:fp,Raffelt:1999tx}. In laboratory or
astrophysical $B$-fields, transitions between these particles and
photons occur~\cite{Sikivie:ip,Raffelt:1987im}, an effect that can
be observed in the laboratory~\cite{pdb}, affects the propagation
of cosmic $\gamma$-rays~\cite{Gorbunov:2001gc,Csaki:2003ef}, and
can modify the apparent brightness of distant astronomical
sources~\cite{Csaki:2001yk,Mortsell:2003ya,Bassett:2003zw}.
%These
%particles also contribute to the magnetically induced vacuum
%bi-refringence, interfering with the corresponding QED
%effect~\cite{Maiani:1986md,Raffelt:1987im}.
Gravitons interact too weakly to be observable in these
situations, while pions are too heavy. However, these effects can
be crucial for new particles, notably the pseudoscalar axions that
arise in the context of the Peccei-Quinn solution to the strong CP
problem and are viable cold dark matter
candidates~\cite{pdb,Bradley:kg}. Galactic axions are currently
being sought by two large-scale Primakoff-type microwave cavity
experiments~\cite{Bradley:kg}.  Anomalous stellar energy loss by
axion emission is constrained by the observed properties of
globular cluster stars, implying \hbox{$g_{a\gamma}\alt
10^{-10}~{\rm GeV}^{-1}$} \cite{Raffelt:1999tx} for the
axion-photon coupling, where the axion-photon interaction is
written in the usual form ${\cal L}_{a\gamma}=
-\frac{1}{4}g_{a\gamma} F_{\mu\nu}\tilde F^{\mu\nu}a
=g_{a\gamma}\,{\bf E}\cdot{\bf B}\,a$. Axions would also
contribute to the magnetically induced vacuum birefringence,
interfering with the corresponding QED
effect~\cite{Maiani:1986md,Raffelt:1987im}. The PVLAS
experiment~\cite{Gastaldi} apparently observes such an effect far
in excess of the QED expectation, although an interpretation in
terms of axion-like particles requires a coupling strength far
larger than existing limits.

On the other hand, the Sun would be a strong axion source and thus
offers a unique opportunity to actually detect these particles by
taking advantage of their back-conversion into X-rays in
laboratory magnetic fields~\cite{Sikivie:ip}.
%The solar axion luminosity due to Primakoff effect is
%$L_a=g_{10}^2\,1.7\times10^{-3}\,L_\odot$, where $g_{10}\equiv
%g_{a\gamma}\,10^{10}~{\rm GeV}$ and
%$L_\odot=3.86\times10^{33}~{\rm erg~s^{-1}}$ is the solar photon
%luminosity.
The expected solar axion flux at the Earth due to the Primakoff
%process is $\Phi_a=g_{10}^2\,3.54\times10^{11}~\rm cm^{-2}~s^{-1}$
process is $\Phi_a=g_{10}^2\,3.67\times10^{11}~\rm cm^{-2}~s^{-1}$
(where $g_{10}\equiv g_{a\gamma}\,10^{10}~{\rm GeV}$) with an
approximate spectrum $d\Phi_a/dE_a=
%g_{10}^2\,4.02\times10^{10}~{\rm cm^{-2}~s^{-1}~keV^{-1}}\,
g_{10}^2\,3.821\times10^{10}~{\rm cm^{-2}~s^{-1}~keV^{-1}}\,
(E_a/{\rm keV})^3/(e^{E_a/1.103~{\rm keV}}-1)$ and an average
energy of 4.2~keV \cite{axionflux}. Axion interactions other than
the two-photon vertex would provide for additional production
channels, but in the most interesting scenarios these channels are
severely constrained, leaving the Primakoff effect as the dominant
one \cite{Raffelt:1999tx}. In any case, it is conservative to use
the Primakoff effect alone when deriving limits on $g_{a\gamma}$.

The conversion probability in a $B$-field in vacuum is
\cite{Sikivie:ip} $P_{a\to\gamma}=(g_{a\gamma}B/q)^2\sin^2(q
L/2)$, where $L$ is the path length and $q=m_a^2/2E_a$
%= 8.45~{\rm cm}^{-1}\,(m_a/{\rm
%eV})^2\,(3~{\rm keV}/E_a)$
is the axion-photon momentum difference. For $q L\alt 1$ where the
axion-photon oscillation length far exceeds $L$ we have
$P_{a\to\gamma}=(g_{a\gamma}B L/2)^2$, implying an X-ray flux of
%\begin{equation}
%\Phi_\gamma=0.74~{\rm cm^{-2}~day^{-1}}\,g_{10}^4\,
%\left(\frac{L}{10~\rm m}\right)^2
%\left(\frac{B}{10~\rm Tesla}\right)^2\!.
%\end{equation}
\begin{equation}
\Phi_\gamma=0.51~{\rm cm^{-2}~d^{-1}}\,g_{10}^4\,
\left(\frac{L}{9.26~\rm m}\right)^2 \left(\frac{B}{9.0~\rm
T}\right)^2\!.
\end{equation}

\noindent For $qL\agt 1$ this rate is reduced due to the
axion-photon momentum mismatch. The presence of a gas would
provide a refractive photon mass $m_\gamma$ so that
$q=|m_\gamma^2-m_a^2|/2E_a$. For $m_a\approx m_\gamma$ the maximum
rate can thus be restored \cite{vanBibber:1988ge}.

The first implementation of the axion helioscope concept was
performed in \cite{Lazarus:1992ry}. More recently, the Tokyo axion
helioscope~\cite{Moriyama:1998kd} of $L= 2.3~{\rm m}$ and
$B=3.9~{\rm T}$ has provided the limit $g_{10}<6.0$ at 95\%~CL for
$m_a\alt 0.03~{\rm eV}$ (vacuum) and $g_{10}<6.8$--10.9 for
$m_a\alt 0.3~{\rm eV}$ (using a variable-pressure buffer
gas)~\cite{Inoue:2002qy}. Limits from crystal
detectors~\cite{Avignone:1997th,Morales:2001we,Bernabei:ny} are
much less restrictive.

{\em CAST Experiment.}---In order to detect solar axions or to
improve the existing limits on $g_{a\gamma}$ an axion helioscope
has been built at CERN by refurbishing a de-commissioned LHC test
magnet \cite{Zioutas:1998cc} which produces a magnetic field of
$B=9.0~\rm T$ in the interior of two parallel pipes of length
$L=9.26~\rm m$ and a cross--sectional area $A=2\times 14.5$
cm$^2$. %The uniformity of the magnetic field along the magnet
%length is better than a 0.5\%.
The aperture of each of the bores
fully covers the potentially axion-emitting solar core
($\sim1/10$th of the solar radius). The magnet is mounted on a
platform with $\pm 8 ^\circ$ vertical movement, allowing for
observation of the Sun for 1.5 h at both sunrise and sunset. The
horizontal range of $\pm 40 ^\circ$ encompasses nearly the full
azimuthal movement of the Sun throughout the year.
%The limitation on the vertical movement
%which comes from
%technical reasons related to the magnet's cryogenic system,
%allows
%us to point to the Sun
%only during about three hours per day on
%average (
%1.5 h at sunrise and 1.5 h at sunset.
The time the Sun is not reachable is devoted to background
measurements.
%In Fig.~\ref{platform} a schematic view
%of the experimental setup is plotted, showing the magnet and the
%platform to move it. The position of the X-ray detectors at both
%ends of the magnet is also marked.
A full cryogenic station is used to cool the superconducting
magnet down to 1.8 K %needed for its superconducting operation
\cite{Barth:2004cx}. The hardware and software of the tracking
system have been precisely calibrated, by means of geometric
survey measurements, in order to orient the magnet to any given
celestial coordinates. The overall CAST pointing precision is better than 0.01$^\circ$ \cite{filming}. %, including all
%sources of inaccuracy such as astronomical calculations, as well
%as spatial position measurements.
%\begin{figure}[t]
%\begin{center}
%\hspace{0 cm} \psfig{figure=platform2.eps,width=90mm}
% \caption{Schematic view of the CAST experimental setup. The 10 m long
% LHC test magnet is mounted on a platform like the one shown in
% the drawing, allowing a movement of
%$\pm 8 ^\circ$ vertically and $\pm 40 ^\circ$ horizontally. The
%detectors are located at both ends of the magnet, exposed to
%axion-induced X-rays during 3 hours per day in average, 1.5 at
%sunrise and 1.5 at sunset.
% \label{platform}}
% \end{center}
%\end{figure}
At both ends of the magnet, three different detectors have
searched for excess X-rays from axion conversion in the magnet
when it was pointing to the Sun. Covering both bores of one of the
magnet's ends, a conventional Time Projection Chamber (TPC) is
looking for X-rays from ``sunset'' axions.
%At the other end, facing "sunrise" axions, a second smaller
%gaseous chamber with novel MICROMEGAS (micromesh gaseous
%structure) \cite{Giomataris:1995fq} readout is placed behind one
%of the magnet bores, while in the other one a Charge Coupled
%Device (CCD) is working in conjunction with a X-ray mirror
%telescope. Both the CCD and the X-ray mirror telescope are
%prototypes developed for X-ray astronomy \cite{richter,jansen}.
%The X-ray mirror telescope produces an "axion image" of the sun by
%focusing the photons from axion conversion, coming parallel to the
%magnet pipe, down to a $\sim$7 mm$^2$ spot on the CCD. The
%signal-to-background ratio should therefore be enhanced by two
%orders of magnitude improving substantially the sensitivity of the
%experiment.
At the other end, facing ``sunrise'' axions, a second smaller
gaseous chamber with novel MICROMEGAS (micromesh gaseous
%structure) \cite{Giomataris:1995fq} readout is placed behind one
% MK change begin
structure -- MM) \cite{Giomataris:1995fq} readout is placed behind
one
% MK change begin
of the magnet bores, while in the other one a focusing X-ray
mirror telescope is working with a Charge Coupled Device (CCD) as
the focal plane detector. Both the CCD and the X-ray telescope are
% MK change begin
prototypes developed for X-ray astronomy \cite{abrixas}.
%\cite{altmann,egle,friedrich}.
% MK change begin
% MK change end
% The X-ray mirror telescope produces an "axion image" of the sun by
% focusing the photons from axion conversion, coming parallel to the
% magnet pipe, down to a $\sim$7 mm$^2$ spot on the CCD. The
% MK change begin
The X-ray mirror telescope can produce an ``axion image'' of the
Sun by focusing the photons from axion conversion to a
$\sim6\,\text{mm}^2$ spot on the CCD. The enhanced
signal-to-background ratio substantially improves the sensitivity
of the experiment. A detailed account of the technical aspects of
the experiment will be given elsewhere.
%\cite{technicalpaper}.

\begin{table*}[t] \centering \footnotesize
\caption{ Data sets included in our result. \label{datasets}}
\begin{tabular}{cccccccc}
\\ \hline\hline
\multicolumn{1}{c}{Data set} & \multicolumn{1}{c}{Tracking
exposure} & \multicolumn{1}{c}{Background exposure} &
\multicolumn{1}{c}{$(g^4_{a\gamma})_{\rm best fit}$ ($\pm 1\sigma$
error)} & \multicolumn{1}{c}{$\chi^2_{\rm null}$/d.o.f} &
\multicolumn{1}{c}{$\chi^2_{\rm min}$/d.o.f} &
\multicolumn{1}{c}{$g_{a\gamma}$(95\%)} \\ & (h) & (h) &
$(10^{-40}$ GeV$^{-4})$ & & & $(10^{-10}$ GeV$^{-1})$\\ \hline
 TPC & 62.7 & 719.9 & $-1.1\pm3.3$ & 18.2/18 & 18.1/17 & $1.55$\\
MM set A & 43.8 & 431.4 & $-1.4\pm 4.5$ & 12.5/14 & 12.4/13 &
$1.67$\\ MM set B & 11.5 & 121.0 & $2.5\pm 8.8$ & 6.2/14 & 6.1/13
& $2.09$
\\ MM set C & 21.8 & 251.0 & $-9.4\pm 6.5$
& 12.8/14 & 10.7/13 & $1.67$
\\ CCD &
121.3 & 1233.5 & $0.4 \pm 1.0$ & 28.6/20 & 28.5/19 & $1.23$\\
\hline
\end{tabular}\end{table*}

{\em Data Analysis and First Results.}--- CAST operated for about
6 months from May to November in 2003, during most of which time
at least one detector was taking data.
% and
%usually all three of them.
%Data
%gathered during this period are normally divided in several data
%sets corresponding to different experimental conditions.
The results presented here were obtained after the analysis of the
data sets listed in Table \ref{datasets}.
%This
%table gathers the available statistics both for Sun-tracking and
%background measurements corresponding to each data set.
%refer to data sets taken with the detectors
%not placed behind the focusing mirror, i. e., the TPC and the
%MICROMEGAS, with the data sets representing the best experimental
%parameters. Data from the TPC detector is in the form of one
%single data set, while for the MICROMEGAS three independent data
%sets are considered. The exposure times of these data sets for Sun
%tracking or background studies are indicated in the table
%\ref{datasets}.
An independent %, though similar,
analysis was performed for each data set. Finally, the results
from all data sets are combined.
%Data from the CCD detector is undergoing a more complex analysis
%profiting from the focusing ability of the mirror telescope.
%Results from this analysis will be presented in a future
%publication.

\begin{figure}[t]
\begin{center}
\psfig{figure=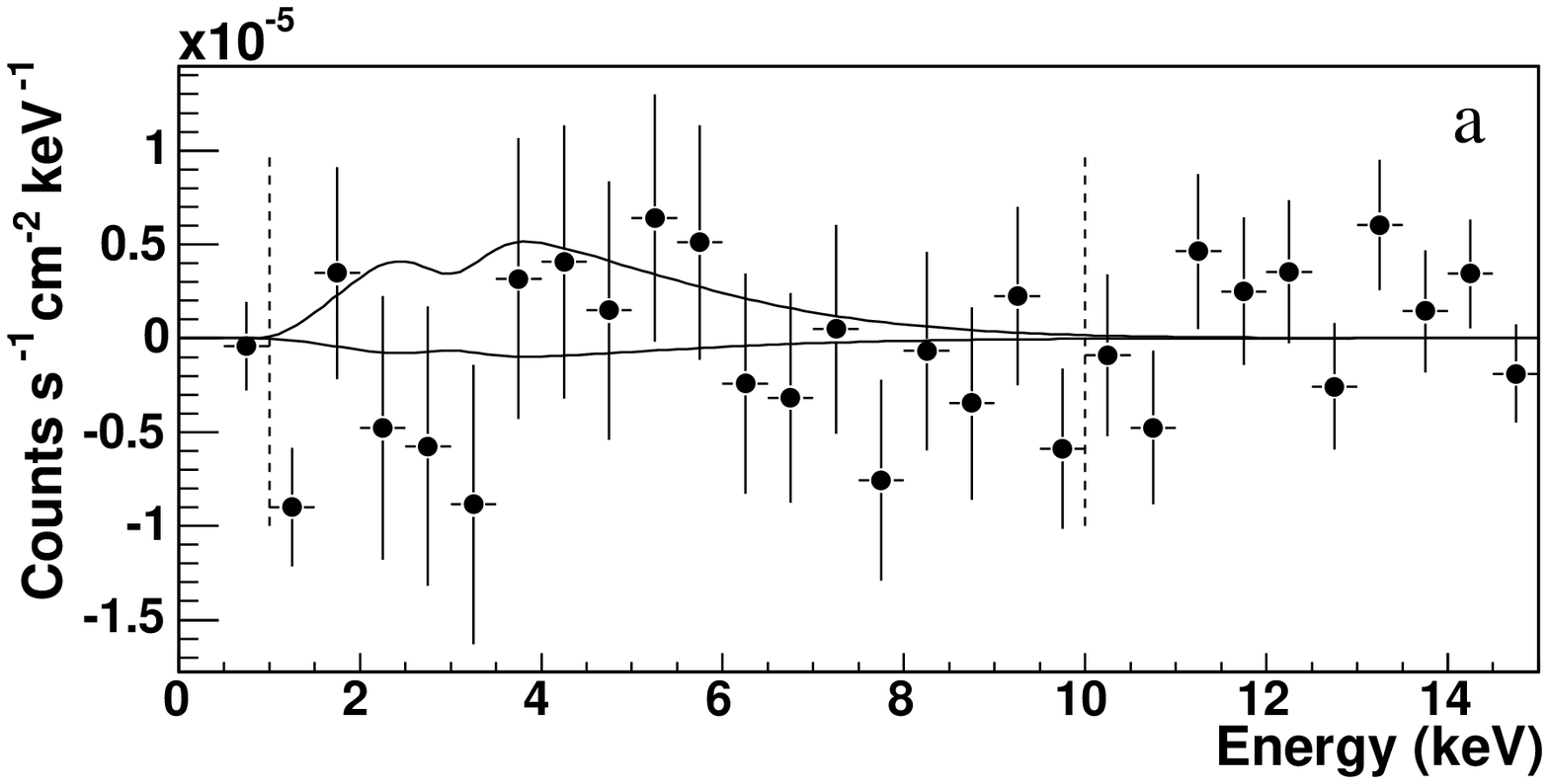,width=75mm}
\psfig{figure=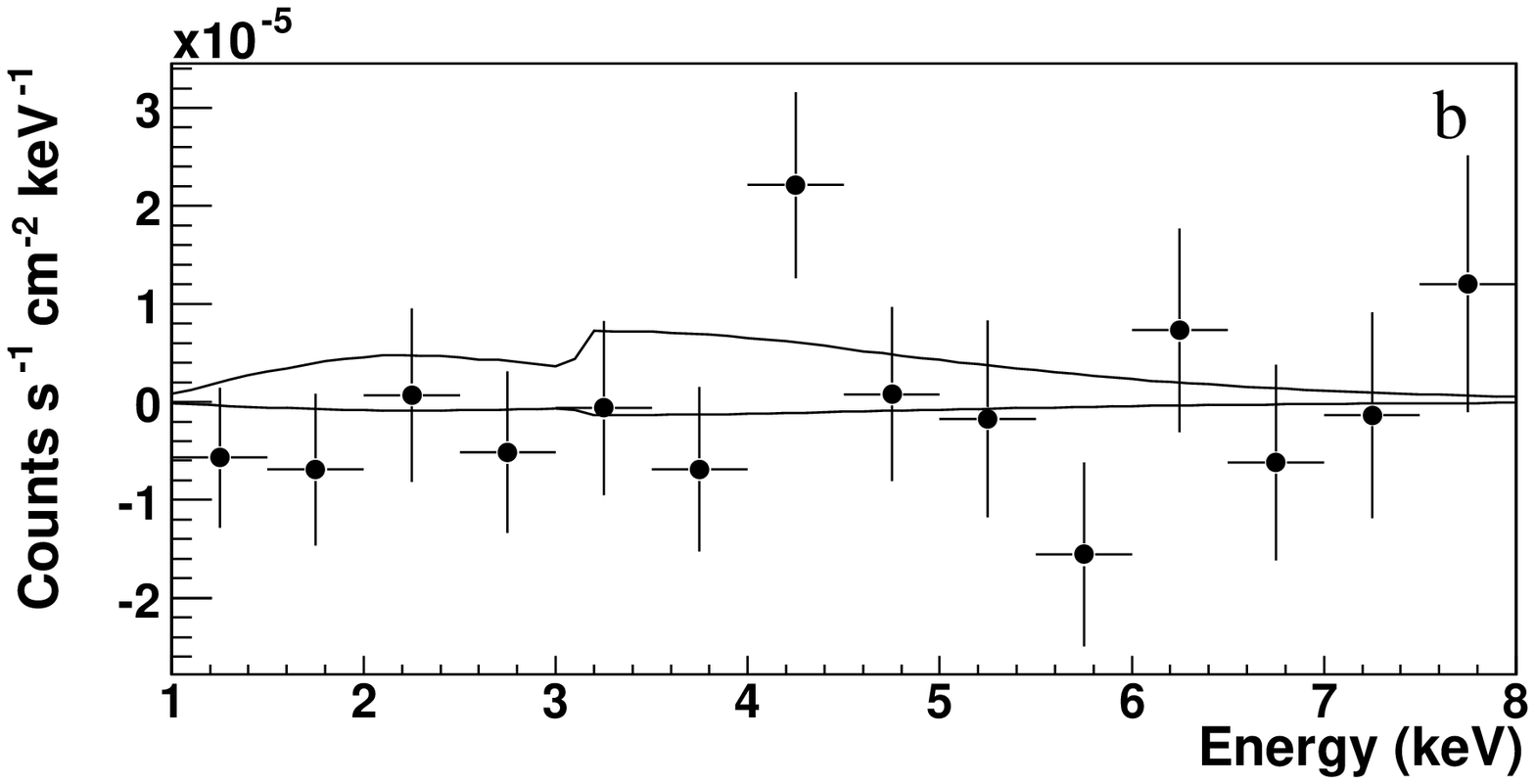,width=75mm}
%\psfig{figure=PlotCCD.eps,width=52mm}
%\vspace{1mm}
\psfig{figure=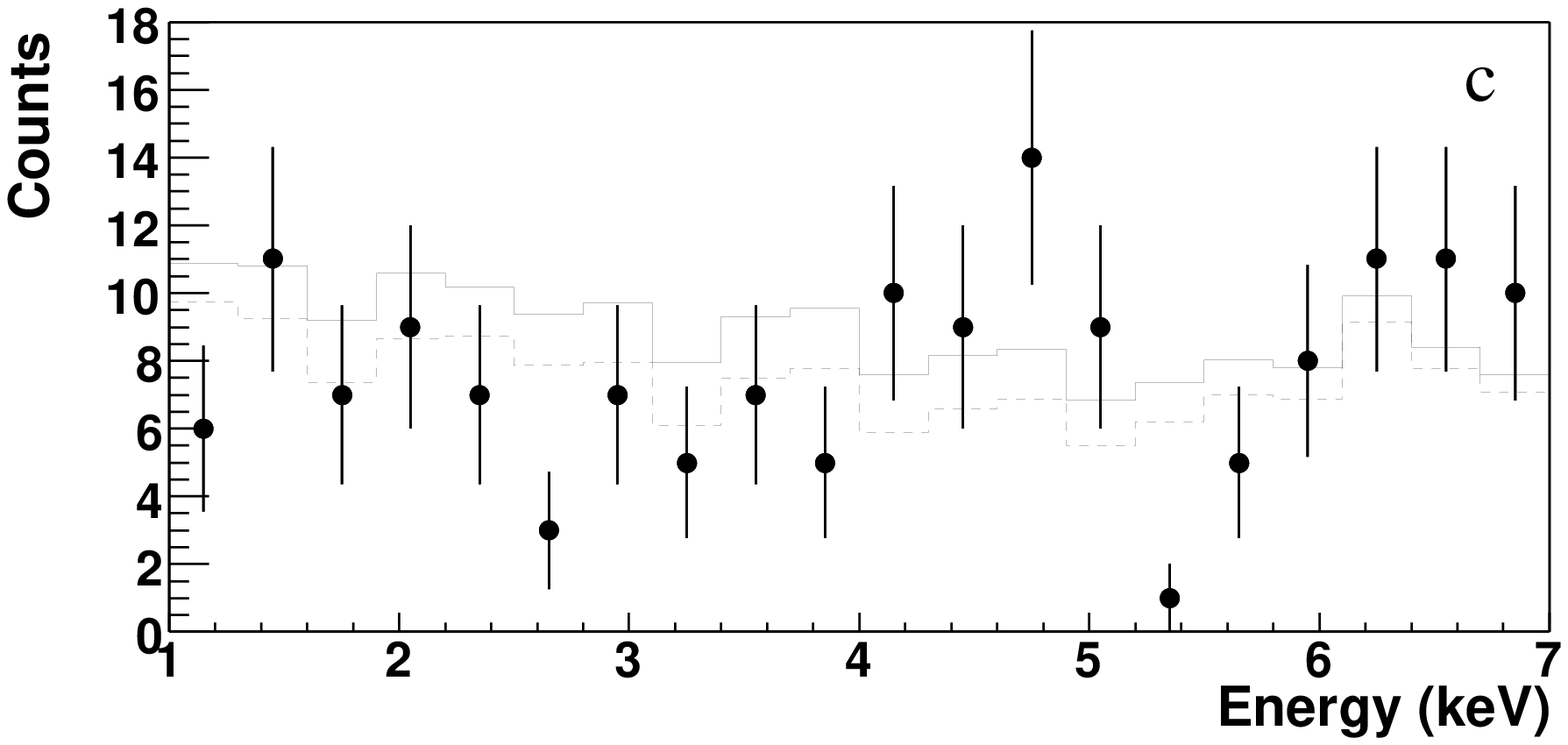,width=75mm}
 \caption{Panels (a) and (b) show respectively the
 experimental subtracted spectrum of the TPC data set and
 MM data set A, together with the expectation for the best
 fit $g_{a\gamma}$ (lower curve) and for the 95\% CL limit on
 $g_{a\gamma}$. For (a) the vertical dashed lines indicate the
 fitting window. The structure at 3 keV in the expected spectrum
 reflects the change in the efficiency curves due to the Ar K-edge
 of the detector gas mixtures. Panel (c) shows both the
 tracking (dots) and background
 (dashed line) spectra of the CCD data set, together with the
 expectation (background plus signal) for $g_{a\gamma}$ at its 95\% CL
 limit, in units of total counts in the restricted CCD area (54.3 mm$^2$) in the
 tracking exposure time (121.3 h).
 \label{subtracted}}
 \end{center}
\end{figure}
An important feature of the CAST data treatment is that the
detector backgrounds are measured with $\sim$10 times longer
exposure during the non-alignment periods.
%, like in the case of TPC and MM, or
%simultaneously with the part of the detector not exposed to the
%X-rays from the magnet, like in the case of the CCD.
%Special effort was made
The use of these data to estimate and subtract the true
experimental background during Sun tracking data is the most
sensitive step in the CAST analysis. To assure the absence of
systematic effects, the main strategy of CAST is the use of three
independent detectors with complementary approaches. In the event
of a positive signal, it should appear consistently, if strong
enough, in each of the three detectors when it is pointing at the
Sun.
%
% To assure that the
%background measurements can be safely used to estimate and
%subtract the true
%experimental background during Sun tracking data, %In particular,
%experimental conditions for background measurements are looked for
%as close as those during the tracking runs.
In addition, an exhaustive recording of experimental parameters
was done, and a search for possible background dependencies on
these parameters was performed.
%Therefore, a possible axion signal is searched for by subtracting
%the properly normalized background spectrum from the one taken
%during solar tracking. In the case of the absence of such a
%signal, the subtracted spectrum should be compatible with zero
%within statistical fluctuations, allowing for the exclusion of the
%axion down to a given value for its coupling with the photon. The
%crucial question that the analysis has to face is whether the
%background measured during non-tracking periods is exactly the
%same as that during tracking. A systematic study of background
%dependencies has been performed in order to answer this question.
A dependence of the TPC background on the magnet position was
found, caused by its relatively large spatial movements at the far
end of the magnet, which resulted in appreciably different
environmental radioactivity levels. Within statistics, no such
effect was observed for the sunrise detectors which undergo a much
more restricted movement. To correct for this systematic effect in
the TPC data analysis, an effective background spectrum is
constructed only from the background data taken in magnet
positions where Sun tracking has been performed and this is
weighted accordingly with the relative exposure of the tracking
data.
%%%%%%%%%%% cut to gain space
%In practice, we distinguish the differential spectrum of the
%complete background data, $dN_b/dE$ from the partial background
%spectra $(dN_b/dE)_i$ corresponding to the data gathered in the
%``cell'' $i$ of the magnet position plane (azimuth-altitude). The
%effective background spectrum to be subtracted from the tracking
%data
%%%%%%%%%%%%%%%%%%%%%%%%%%%
%that will not
%suffer from any systematic effect regarding magnet position and
%therefore could be safely compared with the tracking data is
%defined as
%%%%%%%%%%% cut to gain space
%is therefore built as $(dN_b/dE)_{eff}=\sum_i (dN_b/dE)_i
%\epsilon_i / \sum_i \epsilon_i$, where $\epsilon_i$ is the
%exposure of the tracking data for magnet positions in the cell
%$i$.
%%%%%%%%%%%%%%%%
%The more cells used, the smaller the systematic error will be
%when considering the effective background, but on the other hand
%the higher the statistical errors due to the poor available
%statistics in each of the cells.
%%%%%%%%%%%%%%%% cut to gain space
%For the results presented here, it was proven that a grid of
%3$\times$3 cells is sufficient to reduce this effect well below
%the statistical error in the tracking data.
%%%%%%%%%%%%%%%%%%%%%%%%%%%%%%
%As stated before, for
%the MICROMEGAS data no significant dependence above statistical
%fluctuations has been detected and therefore a straightforward
%subtraction (one single cell) has been performed.
Further checks have been performed in order to exclude any
possible systematic effect. They were based on rebinning the data,
varying the fitting window, splitting the data into subsets and
verifying the null hypothesis test in energy windows or areas of
the detectors where no signal is expected. In general, the
systematic uncertainties are estimated to have an effect of less
than $\sim$10\% in the final upper limits obtained.% quoted later on.
For a fixed $m_a$, the theoretically expected spectrum of
axion-induced photons has been calculated and multiplied by the
detector efficiency curves (determined both by calculation and
measurements).
% of the detectors, including all
%hardware and software efficiency losses, such as window
%transmissions (for TPC and MM), X-ray mirror reflectivity (for
%CCD), detection efficiency and dead time effects.
These spectra, which are proportional to $g_{a\gamma}^4$, are
directly used as fit functions to the experimental subtracted
spectra (tracking minus background) for the TPC and MM. For these
data, the fitting is performed by standard $\chi^2$ minimization.
Regarding the CCD data, the analysis is restricted to the small
area on the CCD where the axion signal is expected after the
focusing of the X-ray telescope. During the data taking period of
2003 a continuous monitoring of the pointing stability
of the X-ray telescope was not yet possible, % \cite{technicalpaper},
therefore a signal area larger than the size of the sun spot had
to be considered. Taking into account all uncertainties of the
telescope alignment, the size of the area containing the signal
was conservatively estimated to be $34\times71\,\text{pixels}$
($54.3\,\text{mm}^2$). As in the other detectors, the background
is defined by the data taken from the same area during the
non-tracking periods but, alternatively, the background in the
signal area was also determined by extrapolating the background
measured during tracking periods in the part of the CCD not
containing the sun spot. Both methods of background selection led
to the same final upper limit on the coupling constant
$g_{a\gamma}$. The resulting low counting statistics in the CCD
required the use of a likelihood function in the minimization
procedure, rather than a $\chi^2$-analysis.
%%%%%%%%%%% cut to gain space
%The likelihood
% function based on a Poissonian PDF is given by:
% \begin{eqnarray}
%   \cal L(\mu) &= {\prod_i^{N} {\mathrm e}^{-\mu_i}\frac{\mu_i^{n_i}}{n_i!}
%     }  \ \ / \ \ {\prod_i^{N} {\mathrm e}^{-n_i}\frac{n_i^{n_i}}{n_i!}}
% \end{eqnarray}
% where $N$ is the number of spectral bins ($20$), $n_i$ the number
% of observed counts in bin $i$, and $\mu_{i}$ the value of the fit
% function in bin $i$. As a fit function $\mu_{i}=s_{i}+b_{i}$ was
% used, where $b_{i}$ is the measured background and $s_{i}$ the
% expected theoretical axion spectrum. The best estimate for
% $g_{a\gamma}^4$ is obtained by minimizing $S=-2\ln{\cal L(\mu)}$.
% In the large sample limit the statistics $S$ is
% $\chi^2$-distributed \cite{pdb}. In the Poissonian regime, but
% with $N$ relatively large, this is usually a reasonable
% assumption. The validity of the $\chi^2$ interpretation of $S$ in
% our particular case, as well as the negligible influence of the
% statistical uncertainty of the background on the final result have
% been verified by Monte Carlo simulations.

The best fit values of $g_{a\gamma}^4$ obtained for each of the
data sets are shown in Table \ref{datasets}, together with their
1$\sigma$ error and the corresponding $\chi^2_{\rm min}$ values
and degrees of freedom. Figure~\ref{subtracted} shows the plots of
some of those fits.
%%%%%%%%%%%%%%%%%%%%%%%%
%%%%%%%%%%%%%%%%%%%%%%%%
%Data of the pn-CCD detector is undergoing a more complex analysis.
%
%Regarding the CCD data, the focusing capability of the X-ray
%telescope allows reducing the background by restricting the
%analysis to the much smaller
%Sun spot area on the CCD chip.% corresponding to the size of the solar
%axion source of $6.3\,\text{mm}^2$.
%%%%%%%%%%%%%%%%%%%%%%%%%%%%%%%%
%%%%%%%%%%%%%%%%%%%%%%%%%%%%%%%%
Each of the data sets is individually compatible with the absence
of any signal as can be seen from the $\chi^2_{\rm null}$ values
shown in Table \ref{datasets}.
%The overall signal strength is proportional to $g_{a\gamma}^4$. In
%figure \ref{subtracted} these theoretical spectra are plotted as
%solid lines for two cases of $g_{a\gamma}^4$, the one giving the
%best fit to the experimental data (lowest "sum of square
%differences", $\chi^2_{min}$, shown in table \ref{datasets}) and
%the one excluded at 95\% CL.
The excluded value of $g_{a\gamma}^4$ was conservatively
calculated by taking the limit encompassing 95\% of the physically
allowed part (i.e. positive signals) of the Bayesian probability
distribution with a flat prior in $g_{a\gamma}^4$. The described
procedures were done using $g_{a\gamma}^4$ instead of
$g_{a\gamma}$ as the minimization and integration parameter
because the signal strength (i.e. number of counts) is
proportional to $g_{a\gamma}^4$.
% and not to the latter.
%>
%>
%>\begin{table} \centering \footnotesize
%>\begin{tabular}{cccc}
%>\\ \hline\hline
%>\multicolumn{1}{c}{Data set} &
%>\multicolumn{1}{c}{$\chi^2_{null}$/d.o.f} &
%>\multicolumn{1}{c}{$\chi^2_{min}$/d.o.f} &
%>\multicolumn{1}{c}{$g_{a\gamma}$(95\%)}\\ \hline
%>
%>\\ TPC & 27.4/28 & 27.3/27 & $1.46\times10^{-10}$ GeV$^{-1}$ \\
%>MICROMEGAS set A & 5.6/12 & 5.1/11 & $1.61\times10^{-10}$
%>GeV$^{-1}$
%>\\ MICROMEGAS set B & 8.5/12 & 8.2/11 & $2.04\times10^{-10}$ GeV$^{-1}$
%>\\ MICROMEGAS set C & 14.0/12 & 10.2/11 & $1.65\times10^{-10}$ GeV$^{-1}$
%>\\
%>\hline
%>
%>\end{tabular}
%>\caption{ Data sets included in the result presented in this
%>letter.\label{chivalues}}
%>\end{table}
The 95\% CL limits on $g_{a\gamma}$ for each of the data sets are
shown in the last column of Table \ref{datasets}. They can be
statistically combined by multiplying the Bayesian probability
functions and repeating the previous process to find the combined
result for the 2003 CAST data:
% (95\% CL) limit:

%\begin{equation}
 $$g_{a\gamma} < 1.16\times 10^{-10}{\rm GeV}^{-1} (95\%{\rm CL}).$$
%\end{equation}

\begin{figure}[!t]
\begin{center}
\psfig{figure=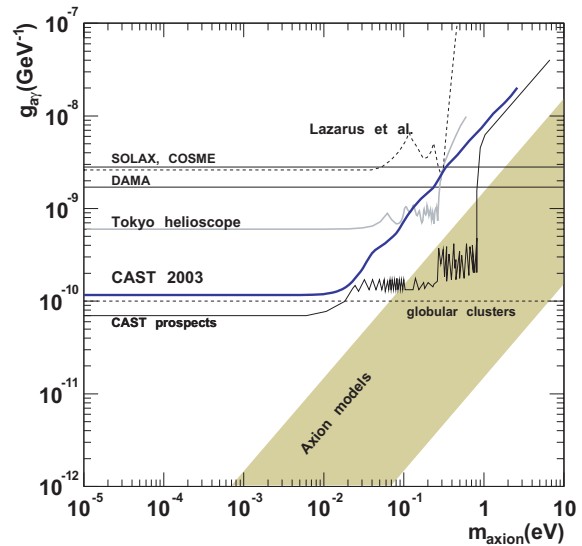,width=75mm}
 \caption{Exclusion limit (95\% CL) from the CAST 2003 data compared with other
 constraints discussed in the introduction. The shaded band represents
 typical theoretical models. Also shown is the future CAST
 sensitivity as foreseen in the
 experiment proposal.
 \label{exclusion}}
 \end{center}
\end{figure}

Thus far our analysis was limited to the mass range
$m_a\alt0.02$~eV where the expected signal is mass-independent
because the axion-photon oscillation length far exceeds the length
of the magnet. For higher $m_a$ the overall signal strength
diminishes rapidly and the spectral shape differs. Our procedure
was repeated for different values of $m_a$ to obtain the entire
95\% CL exclusion line shown in Fig.~\ref{exclusion}.
%
%Regarding the dependence on the axion mass, one has to take into
%account that the expected signal is mass-independent as long as we
%are in the coherence region, i. e., axion masses $\alt$ 0.02 eV.
%The numbers and plots presented are valid for this range of axion
%masses. For higher axion masses the overall signal strength
%diminishes rapidly and the spectral shape differs because of
%coherence patterns. The whole procedure to get the 95\% CL limit
%on $g_{a\gamma}$ as described above has been repeated for
%different values of the axion mass, in order to get the 95\% CL
%complete exclusion line on the ($g_{a\gamma}$--$m_a$) axion
%parameter space. This line is shown in figure \ref{exclusion}
%together with other limits.

{\em Summary.}---Our limit improves the best previous laboratory
constraints \cite{Moriyama:1998kd} on $g_{a\gamma}$ by a factor 5
in our coherence region $m_a\alt0.02$~eV. This result excludes an
important part of the parameter space not excluded by solar age
considerations \cite{Schlattl:1998fz} and is comparable, in this
range of masses, to the limit derived from stellar energy-loss
arguments. A higher sensitivity is expected from the 2004 data
with improved conditions in all detectors, which should allow us
to surpass the astrophysical limit.
%
%Even better results are expected to be achieved from the analysis
%of the data
%%obtained with the mirror telescope as well as the data
%being gathered during 2004 with improved conditions on all
%detectors. They will extent the CAST sensitivity (see fig.
%\ref{exclusion}) beyond the astrophysical limit mentioned above.
In addition, starting in 2005, CAST plans to take data with a
varying-pressure buffer gas in the magnet pipes, in order to
restore coherence for axion masses above 0.02 eV. The extended
sensitivity to higher axion masses will allow us to enter into the
region shown in Fig.~\ref{exclusion} which is especially motivated
by axion models \cite{Kaplan:1985dv}.
%{\em Summary.}--- CAST has taken its first data during 2003.  The
%result of the analysis of these data gives a limit on the
%axion--photon coupling of $1.20\times10^{-10}$ GeV$^{-1}$ for the
%coherence region ($m_a<0.02$ eV). This new limit improves by a
%factor of 5 the previous best laboratory limit of the Tokyo
%helioscope\cite{Moriyama:1998kd} and is comparable to the limit
%obtained by stellar energy-loss arguments.

{\em Acknowledgments.}--- We thank CERN for hosting the experiment
and for the contributions of J.~P.~Bojon, F.~Cataneo,
R.~Campagnolo, G.~Cipolla, F.~Chiusano, M.~Delattre, F.~Formenti,
M.~Genet, J.~N.~Joux, A.~Lippitsch, L.~Musa, R.~De~Oliveira,
A.~Onnela, C.~Rosset and B.~Vullierme. We thank in particular
F.~James for his advice concerning the statistical treatment of
the data. We acknowledge support from NSERC (Canada), MSES
(Croatia), CEA (France), BMBF (Germany), GSRT (Greece),
  RFFR (Russia), CICyT (Spain), and NSF (USA). We dedicate this paper to the memory of our friend and colleague
Angel Morales.

%%%%%%%%%%%%%%%%%%%%%%%%%%%%%%%%%%%%%%%%%%%%%%%%%%%%%%%%%%%%%%%%%%%%%%
%% References %%%%%%%%%%%%%%%%%%%%%%%%%%%%%%%%%%%%%%%%%%%%%%%%%%%%%%%%
%%%%%%%%%%%%%%%%%%%%%%%%%%%%%%%%%%%%%%%%%%%%%%%%%%%%%%%%%%%%%%%%%%%%%%

\end{document}